\documentclass{article}

\usepackage{arxiv}

\usepackage[utf8]{inputenc} 
\usepackage[T1]{fontenc}    
\usepackage{hyperref}       
\usepackage{url}            
\usepackage{booktabs}       
\usepackage{amsfonts}       
\usepackage{amsmath}        
\usepackage{nicefrac}       
\usepackage{microtype}      
\usepackage{lipsum}		
\usepackage{graphicx}
\usepackage{doi}
\usepackage{appendix}
\usepackage{seqsplit}
\usepackage{xcolor}
\usepackage[numbers,sort&compress]{natbib}

\title{New Best-Known Max-Cut Solution for the G63 Instance in the G-Set Benchmark}

\author{ \href{https://orcid.org/0000-0000-0000-0000}{\includegraphics[scale=0.06]{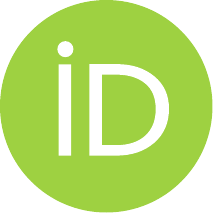}\hspace{1mm}Nikhat Khan}\thanks{Use footnote for providing further
		information about author (webpage, alternative
		address)---\emph{not} for acknowledging funding agencies.} \\
	Electrical and Computer Engineering\\
	University of Virginia\\
	Charlottesville, VA 22903 \\
	\texttt{yqr2yz@virginia.edu} \\
	\And
	\href{https://orcid.org/0000-0000-0000-0000}{\includegraphics[scale=0.06]{orcid.pdf}\hspace{1mm}Nikhil Shukla} \\
	Electrical and Computer Engineering\\
	  University of Virginia\\
	Charlottesville, VA 22903 \\
	\texttt{ns6pf@virginia.edu} \\
}

\begin{document}
\maketitle

\begin{abstract}
	For over two decades, the G-set benchmark has remained a cornerstone challenge for combinatorial optimization solvers. Remarkably, it continues to yield new best-known solutions even to the present day. Here, we report a new best-known Max-Cut of 27,047 for the 7000-node G63 instance—one of the two instances in the benchmark with the largest number of edges. This result is achieved using an optimized Population Annealing Monte Carlo framework, augmented with adaptive control of stochasticity and the periodic introduction of non-local moves, and accelerated on a GPU platform.
    
\end{abstract}


\section{Introduction}
Combinatorial optimization problems (COPs) seek to determine an optimal solution from a discrete set of possible configurations that minimize or maximize a given objective function. These problems underpin a vast range of real-world applications spanning artificial intelligence, biotechnology, finance, to scientific discovery.

Despite their ubiquity, combinatorial optimization remains an unconquered frontier of computing. Many COPs belong to the Non-Deterministic Polynomial-time hard (NP-hard) complexity class, implying that their solution times scale exponentially with problem size in the worst case. Consequently, there is significant motivation to develop new algorithmic and hardware approaches that can advance the current state of the art. This pursuit has inspired vigorous efforts across multiple domains—including quantum, electronic, and optical paradigms—to develop more efficient solvers for these challenging problems, aimed at not only reducing computational time but also improving solution quality.

One of the most studied COPs is the Max-Cut problem, where the goal is to partition the vertices of a graph into two disjoint sets such that the total weight of the edges connecting the two sets is maximized. For unweighted graphs, this objective reduces to finding a partition that maximizes the number of edges crossing between the two sets. The objective function for the problem can be expressed as,
\begin{equation}
\label{eq:maxcut}
\max_{x \in \{-1,+1\}^{N}} \;\; 
\frac{1}{2}\sum_{1 \le i < j \le N} w_{ij}\,\bigl(1 - x_i\,x_j\bigr),
\end{equation}
where, $x_i \in \{-1,+1\}$, $w_{ij}$ represents an edge between nodes $i$ and $j$, and $N$ is the number of nodes.

For the Max-Cut problem, a longstanding cornerstone benchmark for assessing the capabilities and performance of different solvers has been the G-set benchmark, introduced in the early 2000s~\cite{Gset_HelmbergRendl_Stanford}. The benchmark consists of graphs ranging in size from 800 to 20,000 nodes, with edge counts varying between 1,600 to 41,459. These graphs include a variety of topologies—random, toroidal, and planar—and feature both signed and unsigned binary weights. Despite being more than two decades old, the benchmark remains \emph{active}, as new best-known Max-Cut solutions continue to be discovered. Notably, improved Max-Cut solutions for instances such as G72, G77, and G81 were reported as recently as May 2025 by Zick~\cite{zick2025performance}.

Specifically, the 7000-node G63 instance is an unweighted graph that stands out as one of two instances in the G-set (the other being G64) with the largest number of edges (41,459). The previously reported best-known Max-Cut value of 27,045 was achieved using the Global Equilibrium Search (GES) algorithm circa 2015 \cite{shylo2015teams}. To the best of our knowledge, no better cut has been reported for this graph in the nearly ten years since. In this work, we present a new best-known Max-Cut value for this instance, along with the corresponding node configuration.

\section{Background and Prior Work}
As noted earlier, efforts to efficiently solve COPs have progressed along both algorithmic and hardware fronts.

Traditional exact methods—such as the branch-and-bound approach~\cite{RRW10}—guarantee optimal solutions but, as expected, fail to scale efficiently with increasing problem size. This has motivated the use of heuristic and metaheuristic methods that trade strict optimality guarantees for practical speed and scalability.

A widely used baseline for solving the Max-Cut problem is Breakout Local Search (BLS)—an approximate method that alternates steepest-descent improvement with adaptive “breakout” perturbations within an Iterated Local Search loop \cite{BLS}. BLS established or matched many best-known Max-Cut solutions across the G-set benchmark and continues to serve as a strong baseline for comparison.

Subsequently, Global Equilibrium Search (GES)—a classical metaheuristic combining local improvement, path-relinking, and parallel \emph{team} communication among search agents—further advanced the state of the art, setting several new records beyond the BLS era, including the best-known cut for the G63 instance  (27,045) reported circa 2015~\cite{shylo2015teams}. Other algorithmic methods, such as the Multiple Operator Search (MOH) \cite{MOH} have also been used to evaluate the G-set benchmark and offer extended capabilities such as the ability to compute Max-K-Cut ($K\geq2$).

Monte Carlo (MC) methods, such as Simulated Annealing (SA)~\cite{SAmaxcut2015}, have also been extensively explored for solving COPs, but they often suffer from slow convergence and a strong dependence on annealing schedules~\cite{SAlimitations}. More recently, replica-based MC algorithms—including Parallel Tempering (PT)~\cite{tempvariation,huang2024gpu} and Population Annealing Monte Carlo (PAMC)~\cite{machta2010population,PAMC1}—have also been investigated. These approaches typically offer improved mixing, enhanced robustness, and better sampling of complex energy landscapes.

In parallel with these algorithmic advances, a new hardware-centric paradigm has emerged for solving COPs such as the Max-Cut problem, driven by the advent of Ising Machines. A comprehensive review of Ising machines can be found in~\cite{mohseni2022ising}. Ising machines represent a class of physical systems whose intrinsic dynamics evolve toward minimizing the Ising Hamiltonian, thereby providing a novel physical means to solve COPs by mapping them onto equivalent spin systems. As a case in point, computing the Max-Cut of a graph is mathematically equivalent to minimizing the Ising energy of a topologically similar spin graph with antiferromagnetic (negative) coupling~\cite{lucas2014ising}---that is, each positive edge weight in the original graph corresponds to a negative coupling in the spin representation.  

Implementations of Ising machines span multiple physical domains, including quantum, electronic, optical, and acoustic~\cite{litvinenko202550} systems. Examples include quantum annealers (e.g., D-Wave)~\cite{dwavewithembeddedgraphs}; Coherent Ising Machines (CIMs) based on degenerate optical parametric oscillators (DOPOs)~\cite{CIM100kspins} and opto-electronic oscillators~\cite{bohm2019poor}; and electronic implementations employing CMOS oscillators~\cite{antik_accuracy_vs_performance}, spin-torque oscillators~\cite{albertsson2021ultrafast}, and oscillators built using emerging nanoelectronic technologies~\cite{maher2024cmos}. Other related electronic realizations include bistable-latch-based Ising machines~\cite{mallick2023cmos}.  

Furthermore, efforts to realize such systems in physical hardware have been complemented by approaches that emulate their underlying dynamics as algorithms, subsequently accelerating them on high-performance computing platforms such as GPUs and FPGAs. An archetypal example is the Simulated Bifurcation machine (SBM)~\cite{goto2021high,goto2025harnessing} which implements the dynamics of coupled Kerr-nonlinear parametric oscillators. The SBM has been implemented on GPU~\cite{goto2021high} and FPGA platforms~\cite{tatsumura2019fpga} and has helped push the frontiers of performance in solving the G-set benchmark. Examples of other solvers that employ physics-inspired algorithms include those reported in~\cite{Bashar_2024,Yoshimura_2017}. Complementing GPU and FPGA realizations, application-specific integrated circuit (ASIC)-based implementations have also been developed~\cite{CMOSannealing,STATICA_SCA,fujitsudigitalannealer}.

Another computational paradigm being actively explored for solving such problems is probabilistic computing, whose foundations lie in Markov Chain Monte Carlo (MCMC) methods~\cite{camsari2024probabilistic}. Through careful hardware-algorithm co-design and co-optimization, massively parallel Ising machines based on the concept of p-bits (probabilistic bits) have been demonstrated, offering a powerful approach to large-scale combinatorial optimization~\cite{aadit2022massively}. Moreover, such systems can be realized in both special-purpose hardware and conventional digital platforms~\cite{whitehead2024cmos,duffee2025integrated}. 

Nevertheless, as noted earlier, the G-set remains a milestone benchmark for evaluating these methods.

\section{Proposed Approach and Results}
\label{sec:others}
To evaluate the G63 instance, we employ a Population Annealing Monte Carlo (PAMC) algorithm, augmented with adaptive control of stochasticity and the periodic introduction of non-local moves to enhance exploration. The framework is implemented on an NVIDIA RTX A6000 GPU within the University of Virginia's High Performance Computing (HPC) cluster. 

The proposed method achieves a maximum cut value of 27,047, surpassing the previously reported best cut of $27,045$. For runs that produced the best cut, the solver typically ran for 8–24 hours.

A representative phase configuration corresponding to this cut (=27,047) is given by:

\texttt{\seqsplit{
dc7d48547c2371a88a6d797ca52c047414e1e428043a3423d16bfc4b625c86bc87f6fd3fdc10ef3586c6e32cf35340632774a7856ed93a59eeaa8fbcfbe58886d08578a813187d95f47e1827a27b66f24085c14cb0e1d40f49689f4cdce41d4cc01315cebd1e9404ee3b1510a3907bd31d26c65595d8f24151feec32549385377eba7d8745ee11237f9aff0562e406de98d2c36340072e119c29131d5cc004b32be47994d5e08f4c76c11e77988472d278d442f8599e297e00e5ac8166c1dbc965d89909621ed83eec0a8802423ef71867e06fa1de1611fcf38d276cefedd2c87ffce4f12ef3092c3e3f3fe803ce0dae6b1d2b6bc470c08e6d481162c3389cbc4b7c58389a219db96ab9769a2dc4a52e832e9dbc7fa83e58166f3c1b3fb2c4b45896062f55b63b689efd19258a9d18c3e305e1e935c4d2206898d3b38d13dbc33b38c7a5f92e6c20c956a21e84c2351227139c1b50845b53aecabc7c22e1c65337e534fd942af24620f3765f79a4b11da0fe230caff68a060ff1de601529013b8120fc2222a6b4c17545616a4b89bbc8d8f83dc7624395737091f77f58f0de395c09afdbfa66e8824c091fad3eb216acc521a24d2dca6232084a014bb1a1286afb6a76e9a1e84fd1f975f8a723091f3242746f24b99ce66ec32ed329352652ff464c13b1da14612072afe0a36f39d6ca5d5a0586d35517a99369c90544c43016d131074526c59d82874a51bf73d58d1857ce1ec4ffa2f26ca9e0ec557252f862452b8422a79fe4bdcbda006bb720f5308033ba559f181e16244224e2bb83d7b557d0c5b05299cc179134c46e6cf2c494fa5459cb8e631516d9ba81f79ab743426ba0852fb662a0c90c1b0c6527159042b60bb21e1a8353e121ce2c64b30352b1fd0540625013315ab24ceb1d10cdb48f3010c4997bb3ad6445b4e58e8dd7fe84fe35d29611c924a590743ce83584b787a58f415802c919dbbb53a6c9c878f4a77180adec821450c65cca2d3614db9afe641f1959f4fe645889dd59d75e8f6e8e309534a1d10f29498a597d6599468bfcc88e436d72206cc76ded8f1f2f1b2408f49f9536e4be441045d97dd60f4c4225b7ef88e2e0d1dc76a99f52510c059f6d926685d65f14480cbd9a7e89d771ee0aba13cb779e4ea19d7f5068dc53a62f73593217ab83a393e7b1b58d2f9c2594308adbfb69e6cccc86f68e6f7e19210089854636cdb8519d17def0fb}}\\

The above bit-stream has been presented in the Hexadecimal format. To obtain the exact spin configuration: \textbf{(i)} expand the hexadecimal string into binary to obtain binary assignments for variables \(1\)--\(7{,}000\); \textbf{(ii)} map \(\{0,1\}\) to \(\{-1,+1\}\) with \(0 \mapsto -1\).

\section{Conclusion}
For over twenty-five years, many G-set instances have remained formidable challenges for combinatorial optimization solvers. In this work, we report a new best-known Max-Cut solution for the G63 instance in the G-set benchmark, underscoring the continuing potential for advances in COP solver design and performance. A detailed description of the algorithm and its implementation will be made available in the coming days.

\section{Addendum}
The authors recently became aware that Matsubara et al.~\cite{matsubara2020digital} had reported a cut value of 27,052 for the G63 instance in 2020, prior to the publication of our work. To the best of the authors' knowledge, the corresponding configuration was not published. More recently, in 2026, Majchrzak~\cite{majchrzak2026g63}  reported a cut value of 27,050, while Rikito et al.~\cite{ota2026g63} reported a cut value of 27,052. Both reports include the corresponding configurations.

\section*{Acknowledgments}
The authors gratefully acknowledge Prof. Kerem Camsari and Dr. Kenneth Zick for their valuable inputs. This work was supported by the National Science Foundation grant \# 2132198. The authors also acknowledge support from a DAC Analytics Resource Award from the University of Virginia.

\def\bibsection{\section*{References}}  
\bibliography{G63_arXiv_v2}






\end{document}